\documentclass{article} 
\usepackage{iclr2015,times}
\usepackage{hyperref}
\usepackage{url}
\usepackage{graphicx}
\usepackage{amsmath}
\graphicspath{{./Figures/}}

\title{Semantic Modelling with Long-Short-Term Memory for Information Retrieval}

\author{
H. Palangi$^1$, L. Deng$^2$, Y. Shen$^2$, J. Gao$^2$, X. He$^2$,  J. Chen$^2$, X. Song$^2$, R. Ward$^1$\\
$^1$University of British Columbia, Vancouver, BC, Canada\\
$^2$Microsoft Research, Redmond, WA, USA\\
\texttt{$^1$\{hamidp,rababw\}@ece.ubc.ca} \\
\texttt{$^2$\{deng,jfgao,xiaohe,yeshen,jianshuc,xinson\}@microsoft.com} 
}

\begin{document}
\maketitle
\vspace{-0.5cm}
\begin{abstract}
In this paper we address the following problem in web document and information retrieval (IR): How can we use long-term context information to gain better IR performance? Unlike common IR methods that use bag of words representation for queries and documents, we treat them as a sequence of words and use long short term memory (LSTM) to capture contextual dependencies. 
To the best of our knowledge, this is the first time that LSTM is applied to information retrieval tasks. Unlike training traditional LSTMs, the training strategy is different due to the special nature of information retrieval problem. Experimental evaluation on an IR task derived from the Bing web search demonstrates  the ability of the proposed method in addressing both lexical mismatch and long-term context modelling issues, thereby, significantly outperforming  existing state of the art methods for web document retrieval task. 
\end{abstract}
\vspace{-.4cm}
\section{Introduction}
\label{sec:intro}
Two important issues to measure semantic similarities among different text strings include lexical mismatch and the difficulty of incorporating context information.
Lexical mismatch means that one can use different vocabulary items and language styles to express the same concept. This problem is addressed using the translation models \citep{TranslationModels}, the topic models \citep{ref9, ref14}, and the Deep Structured Semantic Model (DSSM)  which makes use of the bag-of-words representation \citep{DSSM}. Incorporation of context information for modelling semantic similarity, on the other hand, can be accomplished by language modelling \citep{ref31, ref11, ref25, ref26}. There are a few recent models which intend to address both issues in a single framework, including the Convolutional DSSM (CLSM) proposed in \citep{CDSSM} and Recurrent DSSM (R-DSSM) proposed in \citep{rdssm}. The main difference between the R-DSSM and CLSM is that while CLSM needs a fixed size sliding window to capture local context information, and a maxpooling layer to capture global context information, the R-DSSM captures both with a recurrent layer without the need for the maxpooling layer.

In this paper, we extend the R-DSSM to incorporate the structure called the Long Short Term Memory DSSM (LSTM-DSSM). The motivations of the extension are as follows. First, due to vanishing and exploding gradient problems, it is difficult for an R-DSSM to capture long term context information effectively. Second, training an R-DSSM is significantly slower than training its DSSM counterpart.  Third, the LSTM-DSSM has the potential to significantly outperform the R-DSSM in practical tasks as evidenced in recent successful applications of the LSTM in large-scale tasks of speech recognition \citep{Sak2014} and machine translation \citep{sequence2sequence}. 

\begin{figure}[t]
\center
\includegraphics[width=0.96\textwidth]{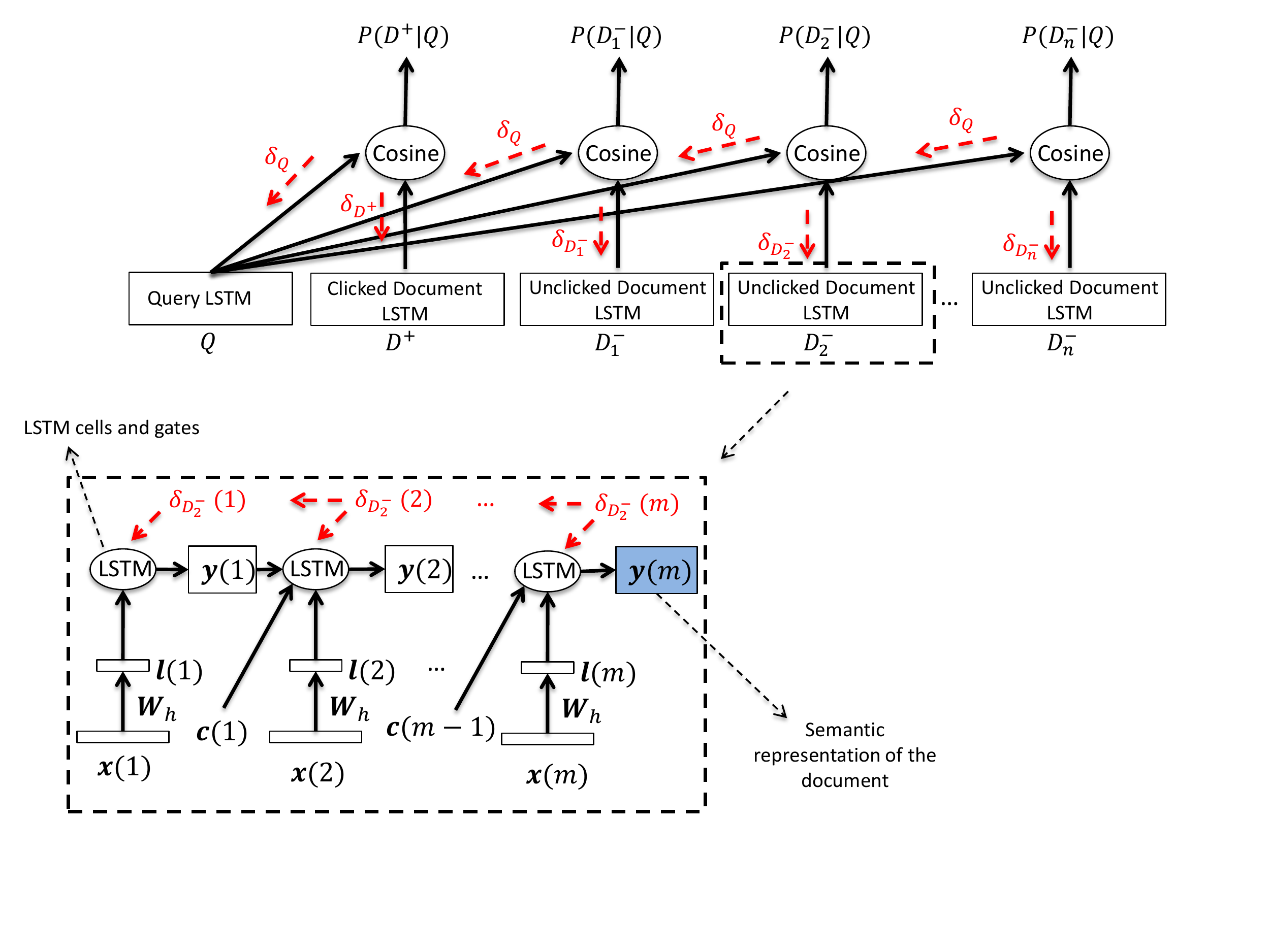}
\vspace{-1.5cm}
\caption{Architecture of the proposed method. $n$ is the number of negative samples (unclicked documents)}
\label{fig:Cosine}
\end{figure}

To the best of our knowledge, this is the first time that LSTM is applied to information retrieval tasks. Unlike training traditional LSTMs, the training strategy is significantly different due to the special nature of information retrieval problem. Specifically, the error signal is generated from the cosine distance between the two semantic embedding vectors of the two text strings (i.e., query and document title), and is then propagated towards the query-LSTM model and the document-LSTM model separately --- see Fig. \ref{fig:Cosine}. From the figure, we also note that, the error signal is only generated from the end of the output sequence and is required to be back propagated to the beginning. This is different from traditional LSTM models, e.g., in speech recognition, where the error signals are generated at every output sample. For this reason, it is critical to use LSTM model in information retrieval problems in order to learn the long-term memory, which is known to be difficult in standard RNN with sigmoid/tanh neurons.

In addition to faster convergence and practical better performance than the R-DSSM, we argue that the LSTM-DSSM can potentially provide valuable information about correlations among different topics and about transitions from one topic to another in a long document.

\vspace{-0.3cm}
\section{The Model}
\label{sec:solution}

\begin{figure}[t]
\center
\includegraphics[width=0.96\textwidth]{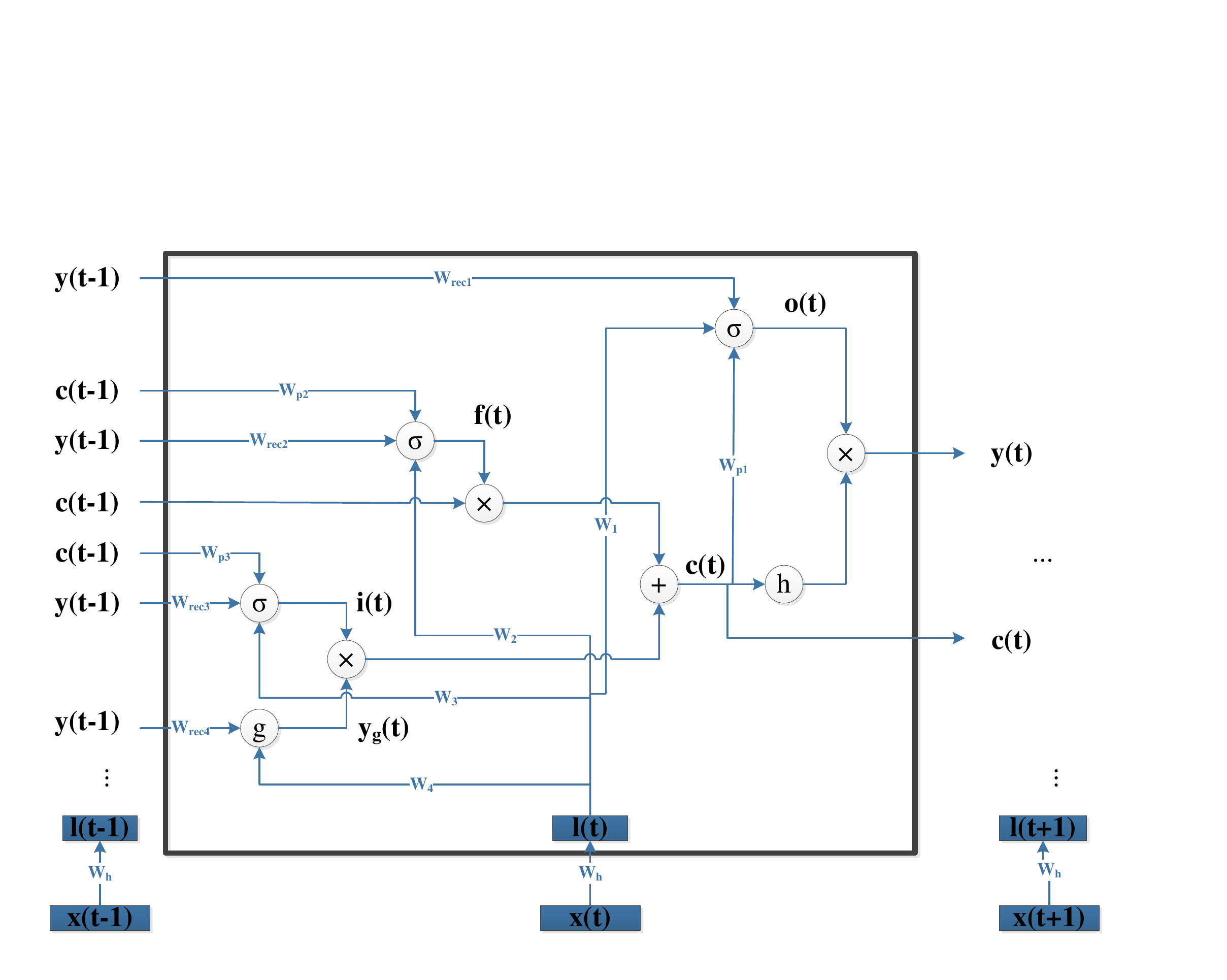}
\caption{The architecture of an LSTM cell used in the LSTM-DSSM}
\label{fig:LSTM Architecture}
\end{figure}

The LSTM-DSSM model developed in this work is aimed to overcome the weakness of the R-DSSM in capturing long-term contextual information effectively. The solution provided by the new model is to incorporate LSTM memory cells, as proposed originally in \cite{lstm} and further developed in \cite{lstm_forget} and \cite{lstm_peephole} by adding forget gate and peephole connections to the architecture. The architecture of the LSTM cell used in the LSTM-DSSM is illustrated in Fig. \ref{fig:LSTM Architecture}, where $\mathbf{i}(t),\;\mathbf{f}(t)\;,\mathbf{o}(t)$ are the input gate, forget gate and output gate, respectively, $\mathbf{c}(t)$ is the cell state, $\mathbf{W}_{p1},\;\mathbf{W}_{p2}$ and $\mathbf{W}_{p3}$ are peephole connections, $g(.)$ and $h(.)$ are $tanh(.)$ functions and $\sigma (.)$ is a sigmoid function. 

We use this architecture to find $\mathbf{y}$ for each word and then use equation \eqref{eq:similarity_func} to find the similarity between query and documents. Subsequently, the LSTM-DSSM is trained using truncated back-propagation-through-time.   

Assuming that we have just one layer of the LSTM (for simplicity of presentation), the mathematical formulation of the LSTM cell according to Fig. \ref{fig:LSTM Architecture} is as follows:
\begin{equation}
\label{eq:yg}
\mathbf{y}_g(t) = g(\mathbf{W}_4\mathbf{l}(t) + \mathbf{W}_{rec4}\mathbf{y}(t-1) + \mathbf{b}_4)
\end{equation}
\begin{equation}
\label{eq:i}
\mathbf{i}(t) = \sigma(\mathbf{W}_3\mathbf{l}(t) + \mathbf{W}_{rec3}\mathbf{y}(t-1) + \mathbf{W}_{p3}\mathbf{c}(t-1) + \mathbf{b}_3)
\end{equation}
\begin{equation}
\label{eq:f}
\mathbf{f}(t) = \sigma(\mathbf{W}_2\mathbf{l}(t) + \mathbf{W}_{rec2}\mathbf{y}(t-1) + \mathbf{W}_{p2}\mathbf{c}(t-1) + \mathbf{b}_2)
\end{equation}
\begin{equation}
\label{eq:c}
\mathbf{c}(t) = \mathbf{f}(t)\circ\mathbf{c}(t-1) + \mathbf{i}(t)\circ\mathbf{y}_g(t)
\end{equation}
\begin{equation}
\label{eq:o}
\mathbf{o}(t) = \sigma(\mathbf{W}_1\mathbf{l}(t) + \mathbf{W}_{rec1}\mathbf{y}(t-1) + \mathbf{W}_{p1}\mathbf{c}(t) + \mathbf{b}_1)
\end{equation}
\begin{equation}
\label{eq:y}
\mathbf{y}(t) = \mathbf{o}(t)\circ h(\mathbf{c}(t))
\end{equation}
where $\mathbf{b}_1,\mathbf{b}_2,\mathbf{b}_3,\mathbf{b}_4$ are bias vectors (not shown in the figure) and $\mathbf{l}(t)$ is the $t$-th word representation after hashing ($\mathbf{W}_h$). 

For training the full LSTM-DSSM, we adopt the cosine similarity measure:
\begin{equation}
\label{eq:similarity_func}
R(Q,D) = \frac{\mathbf{y}_Q(t=T_Q)^T\mathbf{y}_D(t=T_D)}{\Vert\mathbf{y}_Q(t=T_Q)\Vert\Vert\mathbf{y}_D(t=T_D)\Vert}
\end{equation}
where $T_Q$ and $T_D$ are the indexes of the last word in query and document, respectively. The goal is to maximize the likelihood of the clicked document given a query. Therefore, the following optimization problem is to be solved:
\begin{equation}
\label{eq:objective_func}
L(\mathbf{\Lambda})=\underset{\mathbf{\Lambda}}{\operatorname{min}}\;\; -log \prod_{r=1}^{R}P(D_r^+\vert Q_r)
\end{equation}
where $\mathbf{\Lambda}$ is the parameter to be learned and $P(D_r^+\vert Q_r)$ is the probability of clicked document given the $r$-th query as a function of cosine similarity measure according to 
\begin{equation}
\label{eq:L_r}
P(D_r^+\vert Q_r) = \frac{e^{R(Q_r,D_r^+)}}{\sum_{D_{r,j}\in\mathbf{D}}e^{R(Q_r,D_{r,j})}}
\end{equation}
In \eqref{eq:objective_func} and \eqref{eq:L_r}, $Q_r$ is the $r$-th query out of $R$ queries, $D_r^+$ is the clicked document for $r$-th query and $D_{r,j}$ is the $j$-th unclicked document for $r$-th query.
In the learning algorithm, error signals are back propagated through time using following equations which we derived for the LSTM-DSSM:
\begin{equation}
\label{eq:LSTM46}
\mathbf{\delta}_Q^{rec1}(t-1) = [\mathbf{o}_Q(t-1)\circ (1-\mathbf{o}_Q(t-1))\circ h(\mathbf{c}_Q(t-1))]
\circ \mathbf{W}_{rec1}^T.\mathbf{\delta}_Q^{rec1}(t)
\end{equation}
\begin{equation}
\label{eq:LSTM47}
\mathbf{\delta}_Q^{rec_i}(t-1) = [(1-h(\mathbf{c}_Q(t-1)))\circ (1+h(\mathbf{c}_Q(t-1)))
\circ \mathbf{o}_Q(t-1)]\circ \mathbf{W}_{rec_i}^T.\mathbf{\delta}_Q^{rec_i}(t)
\end{equation}
where $Q$ stands for query. And we have derived a similar set of equations for the ``document'' part of the full LSTM-DSSM network.
In the gradient-descent training, we have one large update after folding back in time and adding gradients in each minibatch. We use Nesterov method to accelerate learning convergence. 

\vspace{-0.3cm}
\section{Evaluation Results}

For training and evaluating the LSTM-DSSM and comparing it with the state of the art IR baselines, we have used a real world dataset consisting of 200,000 click-through data collected from Bing search to carry out evaluation experiments. Experimental results are presented in Table \ref{table:Results} using the standard metric of mean Normalized Discounted Cumulative Gain (NDCG) \citep{refNDCG} for evaluating ranking performance. For fair comparisons, we have designed the LSTM-DSSM so that it uses the same number of  model parameters as the baseline R-DSSM and other well known baselines in IR. As is clear from these results, the LSTM-DSSM outperforms all existing baselines significantly in terms of the NDCG metric. Analysis of the results demonstrates the effectiveness of the LSTM cells in capturing long-term correlations in the input text strings, accounting for the success in the information retrieval task designed from Bing search. 
\begin{table}[t]
\caption{Comparisons of NDCG performance measures (the higher the better) of proposed models and a series of baseline models, where {\it nhid} refers to the number of hidden units, {\it ncell} refers to number of cells. The RNN and LSTM-RNN models are chosen to have the same number model parameters as the DSSM and CLSM models: $14.4$M, where $1 \mathrm{M} = 10^6$. The boldface numbers are the best results.}
\label{table:Results}
\begin{center}
\scriptsize
\begin{tabular}{ | c | c | c | c| } 
\hline
Model & NDCG@1 & NDCG@3 & NDCG@10\\ \hline
BM25 & 30.5\% & 32.8\% & 38.8\% \\ \hline
PLSA (T=500) & 30.8\% & 33.7\% & 40.2\% \\ \hline
DSSM (nhid = 288/96), 2 Layers  & 31.0\% & 34.4\% & 41.7\% \\
\hline
CLSM (nhid = 288/96), 2 Layers  & 31.8\% & 35.1\% & 42.6\% \\
\hline
RNN (nhid = 288), 1 Layer  & 31.7\% & 35.0\% & 42.3\% \\
\hline
LSTM-RNN (ncell = 96), 1 Layer & {\bf 33.1\%} & {\bf 36.5\%} & {\bf 43.6\%} \\
\hline
\end{tabular}
\end{center}
\end{table}
\newpage
\bibliography{iclr2015}

\begin{thebibliography}{16}
\providecommand{\natexlab}[1]{#1}
\providecommand{\url}[1]{\texttt{#1}}
\expandafter\ifx\csname urlstyle\endcsname\relax
  \providecommand{\doi}[1]{doi: #1}\else
  \providecommand{\doi}{doi: \begingroup \urlstyle{rm}\Url}\fi

\bibitem[Deerwester et~al.(1990)Deerwester, Dumais, Furnas, Landauer, and
  Harshman]{ref9}
Deerwester, Scott, Dumais, Susan~T., Furnas, George~W., Landauer, Thomas~K.,
  and Harshman, Richard.
\newblock Indexing by latent semantic analysis.
\newblock \emph{Journal of the American Society for Information Science},
  41\penalty0 (6):\penalty0 391--407, 1990.

\bibitem[Gao et~al.(2004)Gao, Nie, Wu, and Cao]{ref11}
Gao, Jianfeng, Nie, Jian-Yun, Wu, Guangyuan, and Cao, Guihong.
\newblock Dependence language model for information retrieval.
\newblock In \emph{Proceedings of the 27th Annual International ACM SIGIR
  Conference on Research and Development in Information Retrieval}, SIGIR '04,
  pp.\  170--177. ACM, 2004.

\bibitem[Gao et~al.(2010)Gao, He, and Nie]{TranslationModels}
Gao, Jianfeng, He, Xiaodong, and Nie, Jian-Yun.
\newblock Clickthrough-based translation models for web search: From word
  models to phrase models.
\newblock In \emph{Proceedings of the 19th ACM International Conference on
  Information and Knowledge Management}, CIKM '10, pp.\  1139--1148, New York,
  NY, USA, 2010. ACM.

\bibitem[Gao et~al.(2014)Gao, Pantel, Gamon, He, Deng, and Shen]{ref14}
Gao, Jianfeng, Pantel, Patrick, Gamon, Michael, He, Xiaodong, Deng, Li, and
  Shen, Yelong.
\newblock Modeling interestingness with deep neural networks.
\newblock \emph{EMNLP}, October 2014.

\bibitem[Gers et~al.(1999)Gers, Schmidhuber, and Cummins]{lstm_forget}
Gers, Felix~A., Schmidhuber, Jürgen, and Cummins, Fred.
\newblock Learning to forget: Continual prediction with lstm.
\newblock \emph{Neural Computation}, 12:\penalty0 2451--2471, 1999.

\bibitem[Gers et~al.(2003)Gers, Schraudolph, and Schmidhuber]{lstm_peephole}
Gers, Felix~A., Schraudolph, Nicol~N., and Schmidhuber, J\"{u}rgen.
\newblock Learning precise timing with lstm recurrent networks.
\newblock \emph{J. Mach. Learn. Res.}, 3:\penalty0 115--143, March 2003.

\bibitem[Hochreiter \& Schmidhuber(1997)Hochreiter and Schmidhuber]{lstm}
Hochreiter, Sepp and Schmidhuber, J\"{u}rgen.
\newblock Long short-term memory.
\newblock \emph{Neural Comput.}, 9\penalty0 (8):\penalty0 1735--1780, November
  1997.

\bibitem[Huang et~al.(2013)Huang, He, Gao, Deng, Acero, and Heck]{DSSM}
Huang, Po-Sen, He, Xiaodong, Gao, Jianfeng, Deng, Li, Acero, Alex, and Heck,
  Larry.
\newblock Learning deep structured semantic models for web search using
  clickthrough data.
\newblock In \emph{Proceedings of the 22Nd ACM International Conference on
  Conference on Information \&\#38; Knowledge Management}, CIKM '13, pp.\
  2333--2338. ACM, 2013.

\bibitem[J\"{a}rvelin \& Kek\"{a}l\"{a}inen(2000)J\"{a}rvelin and
  Kek\"{a}l\"{a}inen]{refNDCG}
J\"{a}rvelin, Kalervo and Kek\"{a}l\"{a}inen, Jaana.
\newblock Ir evaluation methods for retrieving highly relevant documents.
\newblock In \emph{Proceedings of the 23rd Annual International ACM SIGIR
  Conference on Research and Development in Information Retrieval}, SIGIR, pp.\
   41--48. ACM, 2000.

\bibitem[Metzler \& Croft(2005)Metzler and Croft]{ref25}
Metzler, Donald and Croft, W.~Bruce.
\newblock A markov random field model for term dependencies.
\newblock In \emph{Proceedings of the 28th Annual International ACM SIGIR
  Conference on Research and Development in Information Retrieval}, SIGIR '05,
  pp.\  472--479. ACM, 2005.

\bibitem[Metzler \& Croft(2007)Metzler and Croft]{ref26}
Metzler, Donald and Croft, W.~Bruce.
\newblock Latent concept expansion using markov random fields.
\newblock In \emph{Proceedings of the 30th Annual International ACM SIGIR
  Conference on Research and Development in Information Retrieval}, SIGIR '07,
  pp.\  311--318. ACM, 2007.

\bibitem[Palangi et~al.(2015)Palangi, Deng, Shen, Gao, He, Chen, Song, and
  Ward]{rdssm}
Palangi, H., Deng, L., Shen, Y., Gao, J., He, X., Chen, J., Song, X., and Ward,
  R.
\newblock Learning sequential semantic representations of natural language
  using recurrent neural networks.
\newblock In \emph{ICASSP}, 2015.

\bibitem[Platt et~al.(2010)Platt, Toutanova, and Yih]{ref31}
Platt, John~C., Toutanova, Kristina, and Yih, Wen{-}tau.
\newblock Translingual document representations from discriminative
  projections.
\newblock In \emph{Proceedings of the 2010 Conference on Empirical Methods in
  Natural Language Processing, {EMNLP} 2010, 9-11 October 2010, {MIT} Stata
  Center, Massachusetts, USA, {A} meeting of SIGDAT, a Special Interest Group
  of the {ACL}}, pp.\  251--261, 2010.

\bibitem[Sak et~al.(2014)Sak, Senior, and Beaufays]{Sak2014}
Sak, H., Senior, A., and Beaufays, F.
\newblock Long short-term memory recurrent neural network architectures for
  large scale acoustic modeling.
\newblock \emph{INTERSPEECH}, 2014.

\bibitem[Shen et~al.(2014)Shen, He, Gao, Deng, and Mesnil]{CDSSM}
Shen, Yelong, He, Xiaodong, Gao, Jianfeng, Deng, Li, and Mesnil, Gregoire.
\newblock A latent semantic model with convolutional-pooling structure for
  information retrieval.
\newblock In \emph{CIKM}, November 2014.

\bibitem[Sutskever et~al.(2014)Sutskever, Vinyals, and Le]{sequence2sequence}
Sutskever, Ilya, Vinyals, Oriol, and Le, Quoc~V.
\newblock Sequence to sequence learning with neural networks.
\newblock \emph{NIPS}, 2014.

\end{thebibliography}
\bibliographystyle{iclr2015}

\end{document}